# Coverage and metadata completeness and accuracy of African research publications in OpenAlex: A comparative analysis

Patricia Alonso-Álvarez[*] and Nees Jan van Eck [**]

[*] *patricia.alonso@uc3m.es*
https://orcid.org/0000-0002-9305-6024
Universidad Carlos III de Madrid, ROR: https://ror.org/03ths8210, Departamento de Biblioteconomía y Documentación, Laboratorio de Estudios Métricos de la Información (LEMI), Spain
Instituto Interuniversitario de Investigación Avanzada sobre Evaluación de la Ciencia y la Universidad (INAECU), ROR: https://ror.org/004aqbt71, Spain

[**] *ecknjpvan@cwts.leidenuniv.nl*
https://orcid.org/0000-0001-8448-4521
Centre for Science and Technology Studies, Leiden University, The Netherlands

Unlike traditional proprietary data sources such as Scopus and the Web of Science (WoS), OpenAlex emphasizes its comprehensiveness. This study analyzes OpenAlex's coverage and metadata completeness and accuracy of African research publications. To achieve this, OpenAlex is compared with Scopus, WoS, and African Journals Online (AJOL). First, we examine the coverage of African research publications in OpenAlex relative to Scopus, WoS, and AJOL. Then, we assess and compare the availability and accuracy of metadata in OpenAlex, Scopus, and WoS. The findings indicate that OpenAlex offers the most extensive publication coverage. In terms of metadata, OpenAlex provides high coverage for publication and author information, though its coverage of affiliations, references, and funder information is comparatively lower. Metadata accuracy is similarly high for publication and author fields, while affiliation, reference, and funding information show higher rates of missing or incomplete data. Notably, the results demonstrate that both metadata availability and accuracy in OpenAlex improve significantly for publications also indexed in Scopus and WoS. These findings suggest that OpenAlex has the potential to replace proprietary data sources for certain types of analyses. However, for some metadata fields, there remains a trade-off between extensiveness and accuracy.

## 1. Introduction

Bibliographic data sources are essential for bibliometric research, serving as foundational tools for tracking, analyzing, and evaluating scholarly output. For decades, Scopus and Web of Science (WoS) have been the dominant sources of bibliographic data. However, researchers are increasingly exploring new and alternative data sources such as Dimensions, Lens, OpenAIRE, OpenAlex, and Semantic Scholar. These new data sources extend the coverage of traditional data sources by including often overlooked research from diverse regions and languages, thereby enhancing the visibility and use of scientific literature beyond mainstream academic circuits. Moreover, some of these data sources also provide free access and do not impose strict restrictions on the reuse of their bibliographic data. This contributes to increased transparency and reproducibility of bibliometric analyses and is particularly beneficial for researchers in resource-limited settings who may not have access to expensive, proprietary data sources.

Among these new data sources, OpenAlex is regarded by many as particularly promising, due to its extensive coverage and the openness of its data. Launched in January 2022, OpenAlex serves as a free and fully open source of scholarly metadata aimed at enhancing transparency, evaluation, representation, and discovery of research (Priem et al., 2022). OpenAlex data is freely available and can be used and distributed without restrictions. While still in the early stages of its development, OpenAlex has already gained traction in the academic community and has proven to be a promising source for open and reproducible bibliometric analyses. For example, in December 2023, Sorbonne University transitioned from WoS to OpenAlex (Sorbonne University, 2023). In January 2024, the Centre for Science and Technology Studies (CWTS) at Leiden University released the new Leiden Ranking Open Edition based on OpenAlex data (Van Eck, et al., 2024; Waltman, et al., 2024).

Given its growing adoption, a critical evaluation of OpenAlex's strengths and limitations is essential. This study contributes to ongoing research assessing OpenAlex’s utility for bibliometric analysis, with a specific focus on its coverage, metadata availability and completeness, and metadata accuracy of scholarly publications. Given that one of OpenAlex's goals is to enhance inclusivity in scholarly indexing by covering works often excluded from mainstream sources, this paper examines its suitability for studies on traditionally underrepresented regions, with Africa as the case study.

For this analysis, OpenAlex is compared with two major proprietary bibliographic data sources, Scopus and WoS, as well as with African Journals Online (AJOL), a specialized platform that indexes African-based journals to enhance their visibility and impact. While it is worth noting that Scopus and WoS, unlike OpenAlex, aim to be selective rather than comprehensive, comparing OpenAlex against these sources is essential for understanding the opportunities and challenges of using it for bibliometric purposes. Similarly, although AJOL is not a full-fledged bibliographic database like Scopus or WoS, its specific focus on African research makes it a valuable benchmark for assessing OpenAlex's inclusivity and regional representativeness.

This paper has the following three core objectives:

1. To assess the coverage of African research publications in OpenAlex compared to Scopus, WoS, and AJOL.
2. To assess the availability and completeness of metadata in OpenAlex compared to Scopus and WoS.
3. To evaluate the accuracy of OpenAlex metadata compared to Scopus and WoS.

The remainder of this paper is structured as follows. Section 2 presents related work. Section 3 describes the data and methods used to assess the coverage and metadata completeness and accuracy of African research publications in OpenAlex, Scopus, WoS, and AJOL. Section 4 presents the results of the comparative analysis, highlighting key patterns and differences across data sources. Finally, Section 5 concludes the study by summarizing the main findings and discussing their implications, including the relevance for bibliometric research and the further development data sources.

## 2. Related work

### *2.1 Traditional bibliographic data sources: limitations and alternatives*

Traditional bibliographic data sources such as Scopus and WoS have long been regarded as the authoritative sources for identifying relevant publications and journals. They also serve as the primary sources for obtaining data for bibliometric studies and the calculation of bibliometric indicators. However, a growing number of studies have pointed out the biases in these traditional data sources related to the coverage of research fields (Archambault et al., 2006; Larivière & Macaluso, 2011; Lariviérere, Haustein & Mongeon, 2015), regions or countries (Chavarro, 2018; Asubiaro, Onaolapo & Mills, 2024), and languages (van Leeuwen et al., 2001; Lillis & Curry, 2010; Mongeon & Paul-Hus, 2016; Vera-Baceta, Thelwall & Kousha, 2019). Focusing specifically on Africa, Asubiaro et al. (2024) found that journals published in sub-Saharan Africa were the most underrepresented in Scopus and WoS compared to other regions. In a prior study, Asubiaro and Onaolapo (2023) confirmed the limited coverage of traditional bibliographic data sources and highlighted the potential of alternative sources such as Crossref for increased representation.

The impact of these biases ranges from challenges faced by individual authors, such as linguistic exclusion, to broader systemic effects, wherein research from entire regions is marginalized. The exclusion from bibliographic data sources hinders access to relevant local knowledge and affects those who could benefit from it. For example, Moscona and Sastry (2021), in a study on agricultural research, found that Asia and Africa experience a greater productivity loss than other regions due to the adoption of inappropriate technologies. They argue that much of the research in agriculture is tailored to the priorities of high-income countries and is therefore poorly suited to the conditions of other regions. Similar disparities have been observed in other fields, including medicine (Ciarli & Ràfols, 2019; Kumar et al., 2023). Addressing these biases requires not only the production of research that reflects local contexts but also its inclusion and visibility within bibliographic data sources.

The academic community has also highlighted the importance of equitable access to research information. In this regard, the Barcelona Declaration on Open Research Information (2024) recently emphasized the importance of openness of research information (e.g., the metadata of research articles) and the critical role of open scholarly infrastructures. However, scholars advocating for a more diverse scientific system emphasize that openness alone is insufficient and must be combined with a decentralized approach that ensures the richness and diversity of bibliometric data sources (Bambini et al., 2024). This decentralized approach relies on interconnecting global and regional open data sources to promote a federated perspective on research information. An example of the implementation of this approach has successfully connected multiple data sources, such as OpenAlex, SciELO, and PATSTAT, providing public access to the resulting sources, which can then be linked to other open datasets, offering a unique opportunity to access and analyze research information from multiple perspectives (Mazoni & Costas, 2024).

### *2.2 OpenAlex as an alternative bibliographic data source*

Unlike traditional proprietary bibliographic data sources like Scopus and WoS, OpenAlex emphasizes comprehensiveness over selectivity of its content, claiming to offer enhanced coverage of humanities, non-English languages, and the Global South[1]. Moreover, its non-profit nature and the openness of its data provide an opportunity to conduct bibliometric analysis without the restrictions imposed by proprietary data sources. The unrestricted accessibility is particularly significant for researchers in lower-income countries, where accessing proprietary data sources may entail a higher economic burden. Furthermore, OpenAlex's unrestricted data reuse enhances the transparency and replicability of bibliometric analyses.

As a result of its growing popularity, academic research using or examining OpenAlex has increased in recent years. Scholars have used it for a range of studies, including bibliometric analyses (Perianes-Rodríguez, Gómez-Núñez & Olmeda-Gómez, 2024), altmetric research (Mongeon, Bowman & Costas, 2023; Arroyo-Machado & Costas, 2023), and collaboration analyses (Bratt, Langalia & Nonoti, 2023; Okamura, 2024). It has also been used to study open access publishing models, particularly diamond access (Simard et al., 2024), and, due to its extensive regional coverage, North-South inequalities in science production and publishing (Klebel & Ross-Hellauer, 2023; Castro-Torres, 2024).

[1] https://help.openalex.org/hc/en-us/articles/24396686889751-About-us

Another line of research examines the suitability of OpenAlex as a data source for scientometric analysis. Studies focusing on OpenAlex metadata reveal both strengths and limitations. For example, Velez-Estevez et al. (2023) found that the metadata fields included in OpenAlex are generally the same as those in Scopus and WoS. They also highlight the presence of publication and organization identifiers that are absent in Scopus and WoS, such as PMCID, GRID, and ROR ID, which support greater interoperability with other data sources.

However, studies analyzing the completeness and accuracy of OpenAlex's data have identified several limitations. Culbert et al. (2025) found that OpenAlex's reference coverage is comparable to that of Scopus and WoS for the 2015-2022 period, although they raised concerns about the reference computation process. They also found that OpenAlex has better coverage for certain metadata fields, such as ORCIDs, but has lower coverage for others, such as abstracts. Alperin et al. (2024) pointed to limitations in specific fields like language, document type, and citations. In examining the institution field, Zhang et al. (2024) found that missing institutions occur in over 60% of OpenAlex's records, which is particularly problematic in the social sciences and humanities. In a smaller-scale analysis, Delgado-Quirós and Ortega (2024) showed that, although OpenAlex has robust coverage in some fields, primarily those retrieved from Crossref, it inherited some limitations from Microsoft Academic Graph (MAG), resulting in a significant proportion of missing values for certain fields like volume, issue, and pages. Other studies have documented additional errors, inaccuracies, or missing data for specific fields, including document type (Haupka et al., 2024), language (Céspedes et al., 2025), open access (OA) status (Jahn, Haupka & Hobert, 2023), funding acknowledgments (Schares, 2024), and OpenAlex's concepts (Haunschild & Bornmann, 2024).

Given that one of OpenAlex's main strengths is its comprehensiveness, researcher have also investigated its journal and publication coverage in comparison to other bibliographic data sources. An early study revealed that OpenAlex indexes 63.8% of the journals using Open Journal Systems (OJS), compared to only 7.2% in Scopus and only 1.2% in WoS (Khanna et al., 2022). Scholars have also observed that OpenAlex offers better coverage of specific publishing communities, such as gold and diamond OA journals, compared to proprietary data sources, thereby providing valuable insights into its distinctive characteristics (Simard et al., 2024). Research examining OpenAlex's coverage at the publication level further highlights its comprehensiveness. For example, Culbert et al. (2025) confirmed that OpenAlex provides the most extensive coverage of journal articles compared to Scopus and WoS.

### *2.3. Opportunities of inclusive sources: the case of African research*

The development and use of comprehensive bibliographic data sources are particularly relevant for regions that have historically been underrepresented in traditional data sources. Previous research has demonstrated that traditional data sources such as Scopus and WoS do not accurately reflect the global scholarly publication landscape (Khanna et al., 2022). This lack of representativeness disproportionally affects certain regions, most notably the Global South. African research, in particular, remains significantly underrepresented in these traditional data sources. Utilizing publication data from WoS, Tijssen (2007) discovered that African-authored articles accounted for less than 1.5% of all indexed publications between 1980 and 2004 (Tijssen, 2007). More recently, Asubiaro and colleagues (2024) reported that sub-Saharan Africa continues to be the most underrepresented region in both Scopus and WoS.

Although the 1960s marked a period of academic and publishing growth in parts of Africa (Mills et al., 2023), subsequent institutional shifts, particularly the prioritization of publication in so-called 'international' journals, gradually undermined the status of long-established national journals. Commercial publishers, primarily based in the Global North, began to promote global journal indexes and bibliographic databases as markers of quality and legitimacy (Hountondji, 1990). This reinforced a knowledge production model in which inclusion in these sources became synonymous with scholarly value, systematically marginalizing journals from Africa and other regions in the Global South.

In response to the limitations of mainstream bibliographic databases, alternative sources such as African Journals Online (AJOL) have become valuable tools for studying scholarly production in Africa. Established in 1998, AJOL now hosts over 700 journals from 39 African countries, providing a unique view into regional research output. Despite its potential, AJOL remains underutilized due to limited search functionality and a lack of structured metadata. Nonetheless, recent studies have drawn on AJOL to examine the visibility of African journals (Amboka et al., 2024), their representation in global indexes (Alonso-Álvarez, 2024), and patterns of scientific production across countries and disciplines (Boshoff et al., 2024; Asubiaro & Onaolapo, 2023; Ogunfolaji et al., 2022). Still, AJOL and similar sources face their own limitations, including uneven country coverage and the underrepresentation of journals in local languages (Alonso-Álvarez, 2024).

Although recent studies suggest that OpenAlex's journal coverage across countries is relatively balanced, journals located in African countries remain underrepresented compared to their counterparts in the Global North (Chavarro & Alperin, 2024). This underrepresentation may be partly due to factors such as OpenAlex's reliance on Crossref indexing and prioritization of research outputs with a DOI (Chavarro & Alperin, 2024).

To date, most studies examining OpenAlex have focused on its overall coverage and metadata quality, or on specific subsets of publications, without applying a regional lens (Alperin et al., 2024; Culbert et al., 2025; Delgado-Quirós & Ortega, 2024). However, in addition to differences in publication coverage, metadata availability and accuracy may also vary by region, likely following similar patterns of imbalance. Therefore, this paper aims to evaluate both the representation of African research in OpenAlex and the suitability of its metadata for bibliometric analyses, understanding African research as the output of journals located in African countries.

## 3. Data and methods

### *3.1 Data sources*

Our analysis focuses on all publications in African-based academic journals from 1996 to 2022 that are available in OpenAlex, Scopus, WoS, and AJOL. We use only publications in scholarly journals and limit the period to ensure a fair comparison while avoiding biases in the coverage analysis resulting from the specific selection criteria of each data source. For instance, AJOL exclusively indexes journal publications, and Scopus only includes publications from 1996 onward. The versions of the data sources used in the analysis are as follows:

- OpenAlex: We used the OpenAlex snapshot released in November 2023.
- WoS: We used the CWTS in-house version of WoS, updated until September 2023. Our analysis considers the Science Citation Index Expanded (SCIE), the Social Sciences Citation Index (SSCI), the Arts & Humanities Citation Index (AHCI), and the Emerging Sources Citation Index (ESCI).
- Scopus: We used the CWTS in-house version of Scopus, updated in April 2023.

- AJOL: We used data retrieved from the AJOL website in February 2024 using the R libraries ojsr (Becerra, 2022) and rvest (Wickham, 2022).

To identify African-based journals, we selected from each of the above-mentioned data sources all journals associated with an African country and retrieved the ISSNs of those journals. In OpenAlex, we used the country code field available for each source. For Scopus and Web of Science, we relied on the country field of the publisher, as this is the basis on which these databases classify journals geographically. We then constructed a merged journal master list by combining all the retrieved ISSNs to prevent inconsistencies between the geographical classification criteria used by the different data sources. Next, we collected all the publications belonging to the journals in the master list from all four data sources. Table 1 shows the original set of journals retrieved from each data source, along with the number of journals identified when the master list was applied. The substantial increase in journal counts for OpenAlex, Scopus, and WoS when applying the master list confirms that these data sources differ in how they define and classify the geographic origin of journals.

**Table 1. African-based journals as reported by each data source and the final set of journals after using a merged master list**

| | **OpenAlex** | **Scopus** | **WoS** | **AJOL** |
|---|---|---|---|---|
| # African-based journals | 3,115 | 391 | 276 | 739 |
| # journals from master list | 3,511 | 589 | 434 | 739 |

Table 2 presents the number of publications retrieved from each data source after limiting the publication venue to journals and the publication year to the period from 1996 to 2022. Compared to Table 1, the final number of publications in both Scopus and WoS is higher than in AJOL, despite the smaller number of journals. This suggests that journals indexed in Scopus and WoS tend to have higher publication volumes than those indexed in AJOL.

**Table 2. Number of African publications included in each data source**

| | OpenAlex | Scopus | WoS | AJOL |
|---|---|---|---|---|
| # publications | 1,055,096 | 392,625 | 357,879 | 205,029 |

*3.2 Analysis*

The analysis was divided into three steps, examining OpenAlex's coverage of African publications as well as the completeness and accuracy of its metadata:

- Publication coverage analysis: To assess OpenAlex's coverage of African publications, we performed an exact match on DOI across the data sources. OpenAlex was compared with AJOL, Scopus, and WoS to determine its coverage of both indexed and non-indexed African publications.
- Metadata completeness analysis: To assess OpenAlex's metadata completeness, we compared the population of the metadata fields across the three bibliographic data sources: OpenAlex, Scopus, and WoS. AJOL was excluded from this part of the analysis because it is not a bibliographic data source in the sense that it provides access to the metadata of the indexed publications. While some metadata for AJOL publications can be obtained by scraping the website of AJOL, the platform lacks export features comparable to those available in OpenAlex, Scopus, or WoS.

  In the second part of the metadata completeness analysis, we compared the availability of metadata for two subsets of OpenAlex publications: those available in both OpenAlex and either Scopus or WoS (Subset 1), and those exclusively available in OpenAlex (Subset 2). This comparison aimed to assess whether metadata completeness differs between OpenAlex-exclusive publications and those also found in proprietary data sources.
- Metadata accuracy analysis: To evaluate OpenAlex's metadata accuracy, we conducted a manual validation of 60 randomly selected publications. The sample included 30 publications from Subset 1 and 30 publications from Subset 2. This sample size was chosen to allow for meaningful insights while keeping the analysis feasible. For each publication, we compared the metadata in OpenAlex against the information found in the online or PDF version of the publication.

## 4. Results

### *4.1 Publication coverage*

The analysis of OpenAlex's coverage of African publications was carried out using an exact match on DOI. We compared OpenAlex with Scopus, WoS, and AJOL to evaluate its coverage of both indexed and non-indexed African publications. To ensure the reliability of the comparison, we performed DOI deduplication, as some publications in each data source were linked to the same DOI. We excluded these publications from the analysis because it was impossible to determine the correct publication record for each DOI. Table 3 reports the number of African publications without a DOI, with a non-unique DOI, and with a unique DOI for each data source.

**Table 3. Number and percentage of African publications without a DOI, with a non-unique DOI, and with a unique DOI**

| | OpenAlex | | Scopus | | WoS | | AJOL | |
|---|---|---|---|---|---|---|---|---|
| | # | % | # | % | # | % | # | % |
| Publications without a DOI | 163,903 | 16 | 91,181 | 23 | 79,973 | 22 | 109,781 | 54 |
| Publications with a non-unique DOI | 212 | 0 | 952 | 0 | 270 | 0 | 540 | 0 |
| Publications with a unique DOI | 890,981 | 84 | 300,492 | 77 | 277,636 | 78 | 94,708 | 46 |

Figure 1 illustrates the differences in the coverage of African publications between OpenAlex, on the one hand, and Scopus, WoS, and AJOL, on the other hand. As noted, this comparison is limited to publications with a unique DOI in each data source. With 890,981 publications,

OpenAlex is unsurprisingly the most extensive data source, followed by Scopus with 300,492 publications, and WoS with 277,636 articles. AJOL has a smaller coverage due to its high percentage of publications without DOIs (Table 3). Overall, Figure 1 indicates that OpenAlex captures 97% of the publications indexed by Scopus and WoS and nearly all of those by the regional data source AJOL (98%), while also including a significant number of additional publications not covered by the other three sources. These findings suggest that OpenAlex serves as a comprehensive resource for African scholarly publications, particularly those not indexed by traditional proprietary databases. Since all data sources contain publications without DOIs, the accuracy of the comparison could be enhanced by including publications without DOIs, particularly in the case of AJOL.

**Figure 1. Overlap of African publications with a DOI between OpenAlex and Scopus, WoS, and AJOL**

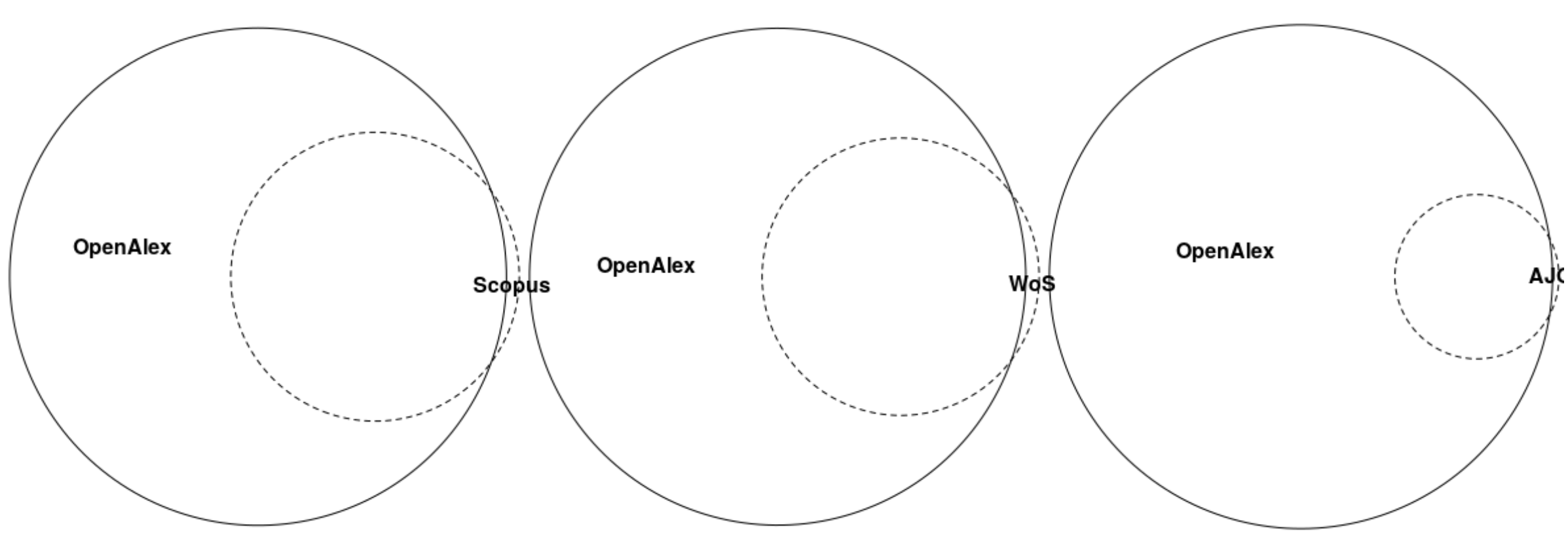

OpenAlex: 890,981
Overlap: 293,436
Scopus: 300,492

OpenAlex: 890,981
Overlap: 270,570
WoS: 277,636

OpenAlex: 890,981
Overlap: 93,225
AJOL: 94,708

*4.2 Metadata completeness: OpenAlex, Scopus and WoS*

To assess metadata completeness of African publications, we examined all publications as reported in Table 2, regardless of whether they had a DOI.

Table 4 presents the percentage of publications for which each metadata field is populated in OpenAlex, Scopus, and WoS. To facilitate readability, the following color-coding scheme is applied: green indicates coverage above 75%, yellow indicates coverage between 50% and 75%, orange indicates coverage between 25% and 50%, and red indicates coverage below 25%. In line with prior research on this topic (Velez-Estevez et al., 2023; Singh & Singh, 2023), metadata fields are grouped into several dimensions (i.e., publication information, author information, affiliation and institution information, funding information, reference information) to enable a structured comparison across data sources.

**Table 4. Metadata completeness of African research publications in OpenAlex, Scopus, and WoS**

| Metadata field | OpenAlex | | Scopus | | WoS | |
|---|---|---|---|---|---|---|
| | # | % | # | % | # | % |
| Publications | 1,055,096 | - | 392,625 | - | 357,879 | - |
| Publication information | | | | | | |
| *Identifiers* | | | | | | |
| DOI | 891,193 | 85 | 301,444 | 77 | 277,906 | 78 |
| *Bibliographic data* | | | | | | |
| Date | 1,055,096 | 100 | 157,851 | 40 | 240,530 | 67 |
| Year | 1,055,096 | 100 | 392,625 | 100 | 357,879 | 100 |
| Volume | 923,477 | 88 | 387,230 | 99 | 323,301 | 90 |
| Issue | 776,954 | 74 | 257,196 | 66 | 214,443 | 60 |
| Article number | - | - | 143,664 | 37 | 142,216 | 40 |
| First page | 786,080 | 75 | 247,126 | 63 | 211,341 | 59 |
| Last page | 776,995 | 74 | 234,064 | 60 | 211,341 | 59 |
| *OA status* | | | | | | |
| OA status | 1,055,096 | 100 | 292,673 | 75 | 357,879 | 100 |
| *Content* | | | | | | |
| Title | 1,047,394 | 99 | 392,623 | 100 | 357,879 | 100 |
| Abstract | 763,341 | 72 | 353,930 | 90 | 311,889 | 87 |
| Author keywords | - | - | 193,651 | 49 | 176,619 | 49 |
| Generated keywords | 876,803 | 83 | - | - | 233,819 | 65 |
| *Other* | | | | | | |
| Type | 1,055,096 | 100 | 392,625 | 100 | 357,879 | 100 |
| Language | 1,032,565 | 98 | 392,621 | 100 | 357,879 | 100 |
| Author information | | | | | | |
| Author names | 1,009,860 | 96 | 388,584 | 99 | 354,973 | 99 |
| ORCIDs | 463,846 | 44 | 85,168 | 22 | 35,765 | 10 |
| Affiliation and institution information | | | | | | |
| Affiliations | 639,027 | 61 | 374,451 | 95 | 339,386 | 95 |
| Unified institutions | 509,518 | 49 | 371,892 | 95 | 309,265 | 86 |
| ROR IDs | 509,518 | 49 | - | - | - | - |
| Reference information | | | | | | |
| References | 474,385 | 45 | 348,936 | 89 | 338,662 | 95 |
| Funding information | | | | | | |
| Funding text | - | - | 65,207 | 17 | 117,218 | 33 |
| Funders | 65,795 | 6 | 63,079 | 16 | 118,224 | 33 |
| ROR IDs | 58,264 | 6 | - | - | - | - |

Publication information: All three data sources demonstrate good DOI coverage, with OpenAlex displaying a slightly higher percentage of publications with DOIs. Other identifiers were not included in the comparison due to their limited use across disciplines (e.g., PMID, PMCID, arXiv ID) or because they are specific to certain data sources (e.g., MAG ID, which appears only in OpenAlex as a result of its use of MAG data). OpenAlex demonstrates the highest overall coverage of bibliographic data. WoS has the lowest coverage for the issue field, while Scopus shows notably low coverage for the date field. For the page and article number fields, it is important to note that they are complementary. This is because Scopus and WoS provide either page numbers or article numbers, depending on the publication format. When both fields are considered together, coverage increases to nearly 100% for Scopus and 99% for WoS. In contrast, OpenAlex typically reports article numbers in the page fields.

Regarding OA information, OpenAlex and WoS include the OA status for all their publications. Scopus, however, provides this information for slightly less than 75% of its publications.

In terms of content-related information, a title is available for nearly all publications across the data sources, with OpenAlex missing a title for 1% of its publications. Scopus and WoS also have substantial coverage of abstracts, while OpenAlex has a lower coverage of 72%. Unlike

Scopus and WoS, OpenAlex does not include author keywords. However, OpenAlex generates keywords for over 80% of its publications. In contrast, WoS provides database generated keywords for only 65% of its publications, and Scopus does not include such keywords at all.

Nearly all publications in the three data sources include information about their type and language. By "type" we mean the category or classification of the publication—such as article, review, editorial, or conference paper—that describes the nature of the publication. It is important to note that the publication type information in Scopus and WoS is more detailed than in OpenAlex.

Author information: Most publications in all three data sources include author names, with coverage consistently above 95%. OpenAlex shows a slightly lower coverage (96%) compared to Scopus and WoS (both 99%). In terms of author identifiers (ORCID), coverage remains below 50% across all data sources. Interestingly, the availability of ORCIDs is significantly higher in OpenAlex (44%) than in Scopus (22%) and WoS (10%).

Affiliation and institution information: All three data sources aim to make the (raw) affiliation strings provided in publications available and link them to standardized, unified institutions using their own internal registry. Scopus uses Affiliation IDs, WoS employs Organization-Enhanced, and OpenAlex relies on a list of institutions that is largely based on the Research Organization Registry (ROR).

Most publications in Scopus and WoS include affiliation and institution data. OpenAlex includes affiliation data for 61% of its publications and institution data for 49%. Importantly, OpenAlex is the only source that provides standardized institution identifiers (ROR IDs) for publications with affiliations, something Scopus and WoS currently lack.

Reference information: The coverage of references is highest in Scopus and WoS, with 89% and 95% of publications including references, respectively. OpenAlex lags significantly behind with only 45% of its publications containing reference data. This lower coverage is partly due to OpenAlex including only linked references (i.e., those that point to publications also included in OpenAlex). Furthermore, many publications in OpenAlex are missing complete reference lists, which further contributes to the lower coverage. It is also important to note that while Scopus and WoS provide detailed reference information, OpenAlex does not. Instead, OpenAlex only indicates the existence of references to other indexed publications, without including any additional reference details.

Funding information: Funding information is limited in all three data sources. WoS offers the highest number of publications with funding organizations and, like Scopus, provides the full text of acknowledgements. OpenAlex, however, does not contain acknowledgement texts. While OpenAlex does offer ROR IDs for funding organizations, its coverage remains low at just 6% of its publications.

### *4.3 Metadata completeness: OpenAlex subsets*

To further assess the availability and completeness of OpenAlex metadata, we compared two subsets of publications: those indexed in both OpenAlex and either Scopus or WoS (Subset 1), and those found exclusively in OpenAlex (Subset 2). As shown in Table 5, Subset 1 consistently exhibits higher metadata availability across all fields compared to Subset 2. This suggests that publications also included in proprietary databases tend to have more complete metadata in OpenAlex. When comparing the results of Subset 1 in Table 5 to those in Table 4, we observe that OpenAlex's metadata availability is comparable to that of Scopus and WoS. It is important to note, however, that Table 4 includes all publications regardless of DOI presence, while Table 5 only includes publications with DOIs.

**Table 5. Metadata completeness of African research publications in OpenAlex subsets**

| Metadata field | *Subset 1:* Publications in OpenAlex and Scopus or WoS | | *Subset 2:* Publications only in OpenAlex | |
|---|---|---|---|---|
| | # | % | # | % |
| Publications | 306,216 | - | 584,977 | - |
| Publication information | | | | |
| *Identifiers* | | | | |
| DOI | 306,216 | - | 584,977 | - |
| *Bibliographic data* | | | | |
| Date | 306,216 | 100 | 584,977 | 100 |
| Year | 306,216 | 100 | 584,977 | 100 |
| Volume | 286,055 | 93 | 490,645 | 84 |
| Issue | 166,085 | 54 | 475,787 | 81 |
| Article number | - | - | - | - |
| First page | 233,743 | 76 | 417,335 | 71 |
| Last page | 233,685 | 76 | 415,161 | 71 |
| *OA status* | | | | |
| OA status | 306,216 | 100 | 584,977 | 100 |
| *Content* | | | | |
| Title | 306,210 | 100 | 577,283 | 99 |
| Abstract | 285,668 | 93 | 360,395 | 62 |
| Author keywords | - | - | - | - |
| Generated keywords | 302,027 | 99 | 425,415 | 73 |
| *Other* | | | | |
| Type | 306,216 | 100 | 584,977 | 100 |
| Language | 304,610 | 100 | 565,343 | 97 |
| Author information | | | | |
| Author names | 303,891 | 99 | 542,083 | 93 |
| ORCIDs | 237,952 | 78 | 171,407 | 30 |
| Affiliation and institution information | | | | |
| Affiliations | 258,323 | 84 | 320,852 | 55 |
| Unified institutions | 250,981 | 82 | 214,332 | 37 |
| ROR IDs | 250,981 | 82 | 214,332 | 37 |
| Reference information | | | | |
| References | 259,487 | 85 | 178,987 | 31 |
| Funding information | | | | |
| Funding text | - | - | - | - |
| Funders | 60,443 | 20 | 5,933 | 1 |
| ROR IDs | 53,125 | 17 | 5,686 | 1 |

Publication information: The date and year fields show complete coverage across both subsets of OpenAlex publications. The volume field is more complete in Subset 1 (93%) than in Subset 2 (84%). Interestingly, the issue field shows notably higher coverage in Subset 2 (81%) compared to Subset 1 (54%). The first page and last page fields show comparable levels of completeness across both subsets, at 76% and 71%, respectively. The OA status has full coverage at 100% in both subsets.

Regarding content information, title coverage is high in both subsets (99-100%). However, abstract availability is notably higher in Subset 1 (93%) than in Subset 2 (62%). Similarly, database generated keywords are available for 99% of publications in Subset 1, compared to 73% in subset 2.

Author information: Author name information is nearly complete in Subset 1 (99%) but less so in Subset 2 (93%). This discrepancy is particularly concerning, as author names are a core component of bibliographic metadata. There are also significant differences in ORCID coverage, with 78% coverage in Subset 1 compared to just 30% in Subset 2.

Affiliation and institution information: The availability of affiliations is significantly higher in Subset 1 (84%) compared to Subset 2 (55%). Unified institutions and ROR IDs follow a similar pattern, with 82% coverage in Subset 1 and only 37% in Subset 2.

Reference information: The reference field shows one of the most significant differences between the subsets, with 85% coverage in Subset 1 compared to just 31% in Subset 2.

Funding information: The availability of funding organizations is limited in both subsets. In Subset 1, funder coverage reaches 20%, while in Subset 2 it is especially low at only 1%.

*4.4 Metadata accuracy: publications in OpenAlex, Scopus, and WoS*

This section presents the results of a manual analysis of 30 randomly selected African publications indexed in OpenAlex, Scopus, and WoS (from Subset 1) to evaluate the accuracy of OpenAlex metadata and compare it with the other two data sources. Following the framework of Zhang et al. (2024), publications are classified into four categories based on the availability and accuracy of information in the metadata fields:

- Complete missing information (CMI): All the information in the field is missing. For example, none of the publication's references are provided.
- Partially missing information (PMI): Some, but not all, of the expected values in a multi-valued metadata field are missing. For example, a publication lists three institutions, but only two are reported.
- Non-matching information (NMI): Information in the field differs from what is available in the online or PDF version of the publication. For example, discrepancies in author names, non-matching page numbers, or fields with extra or incomplete information.
- Full information (FI): All information in the field is available and matches exactly with the information in the online or PDF version of the publication.

Table 6 presents the results of this accuracy analysis for OpenAlex (for clarity, the results for Scopus and WoS are not included in the table but are discussed in the text). In the table, the number of missing or non-matching elements is indicated in parentheses. For example, under NMI for authors, seven authors from five publications are misspelled.

**Table 6. Accuracy of metadata of African publications indexed in OpenAlex, Scopus, and WoS based on a sample (n = 30) of publications**

| Metadata field | CMI | PMI | NMI | FI |
|---|---|---|---|---|

|  | # | % | # | % | # | % | # | % |
|---|---|---|---|---|---|---|---|---|
| Publication information | | | | | | | | |
| DOI | 0 | 0 | - | - | 0 | 0 | 30 | 100 |
| Publication year | 0 | 0 | - | - | 0 | 0 | 30 | 100 |
| Volume | 3 | 10 | - | - | 0 | 0 | 27 | 90 |
| Issue | 18 | 60 | - | - | 1 | 3 | 11 | 37 |
| Article number | - | - | - | - | - | - | - | - |
| First page | 1 | 3 | - | - | 15 | 50 | 14 | 46 |
| Last page | 1 | 3 | - | - | 15 | 50 | 14 | 46 |
| OA status | 0 | 0 | - | - | 2 | 6 | 28 | 92 |
| Title | 0 | 0 | - | - | 0 | 0 | 30 | 100 |
| Abstract | 1 | 3 | - | - | 1 | 3 | 28 | 94 |
| Type | 0 | 0 | - | - | 2 | 6 | 28 | 94 |
| Language | 0 | 0 | - | - | 0 | 0 | 30 | 100 |
| Author information | | | | | | | | |
| Author names | 0 | 0 | 0 | 0 | 5 (7) | 23 | 25 | 83 |
| Author order | 0 | 0 | 0 | 0 | 0 | 0 | 30 | 100 |
| Corresponding author | 0 | 0 | 0 | 0 | 6 (6) | 20 | 24 | 83 |
| Affiliation and institution information | | | | | | | | |
| Raw affiliation strings | 0 | 0 | 0 | 0 | 0 | 0 | 30 | 100 |
| Institutions | 0 | 0 | 6 (7) | 20 | 0 | 0 | 24 | 80 |
| Reference information | | | | | | | | |
| References | 1 | 3 | 25 (283) | 83 | 0 | 0 | 4 | 13 |
| Funding information | | | | | | | | |
| Funders | 4 | 13 | 2 (2) | 6 | 0 | 0 | 24 | 80 |
| Source information | | | | | | | | |
| Journal title | - | - | - | - | 0 | 0 | 30 | 100 |
| e-ISSN | 0 | 0 | 0 | 0 | 0 | 0 | 30 | 100 |
| p-ISSN | 0 | 0 | 0 | 0 | 0 | 0 | 30 | 100 |

Publication information: Most metadata fields related to publication information in OpenAlex are accurate and consistent with the online or PDF version of the publication. The DOI, publication year, and title fields are correct for all 30 publications. The volume, OA status, and abstract fields also show high accuracy rates (90%, 92%, and 94%, respectively), with only a few missing or non-matching values. However, the issue and page fields present some notable discrepancies.

In the case of the issue field, the high NMI percentage (60%) is entirely due to journals that publish only one issue per year. In such cases, OpenAlex often omits the issue number and leaves it blank, while journals typically label the single issue as 'Issue 1'.

Regarding the page fields, 50% of the publications present NMI. This is mainly due to OpenAlex assigning page numbers to articles in journals that use article numbers instead. Unlike other data sources such as WoS and Scopus, OpenAlex does not include an article number field. When journals use article numbers instead of page numbers, they may still paginate their articles internally from 1 to the total number of pages, pagination that is not related to the order within a journal issue. This kind of pagination is useful for locating information within the article itself but lacks bibliometric value. OpenAlex typically includes this internal pagination in the page fields. Consequently, although the page fields show high coverage, they provide different information than traditional page metadata or article numbers and cannot be used to adequately identify publications.

Scopus and WoS similarly tend to omit the issue number when journals have only one issue per year. The other fields are mostly accurate, with minor exceptions. For instance, in both data source three publications show NMI in the OA status field (two of them belonging to the green OA route), and one publication has NMI in the document type field.

<u>Author information</u>: OpenAlex shows a high degree of accuracy. It correctly lists all authors for the 30 publications in the sample. However, the names of seven authors, mostly Chinese, are misspelled.

ORCIDs are not included in Table 6 because OpenAlex does not retrieve them from the publication itself but from its internal author entities, which rely heavily on information from ORCID. As a result, the availability and content may differ from what is found in the online or PDF version of the publication. However, all ORCIDs recorded by OpenAlex have been compared with the information in the publications and verified using the ORCID website. OpenAlex includes ORCID data for 62% of the authors in the publication sample. In only four cases does OpenAlex omit an ORCID that is found in the online or PDF version of the publication. Over half of the ORCIDs reported by OpenAlex are not found in the publication itself. As noted, this discrepancy is due to OpenAlex's internal structure and retrieval process. All but one of the ORCIDs recorded in OpenAlex are accurate and correspond to the publication's author. Two ORCIDs relate to OpenAlex author profiles with the same name as the author but do not match the ORCID in the publication. Since those author profiles display no additional information beyond the name on the ORCID website, it remains unclear whether they represent duplicate profiles for the same person or entirely different individuals with the same name.

We also reviewed the order of authors and the corresponding author field. The order of authors is accurate in all cases. However, six publications lack corresponding author information.

In Scopus and WoS, names are mostly correct, with one exception in WoS where there is a mix-up between first and last names. ORCID coverage is more limited in these data sources. However, Scopus includes all ORCIDs listed in the online or PDF version of the publications, while WoS omits some. The ORCIDs that are present in both data sources are accurate and correspond to the correct persons. The order of the authors is also correct in all cases.

<u>Institution information</u>: The results show differences between the raw affiliation and institution fields. All but one publication have correct and complete raw affiliation strings. However, six publications have at least one missing institution in the institution field. OpenAlex also disaggregates some institutions. For example, recording a hospital and its parent university separately.

As with ORCIDs, ROR IDs are not retrieved from the publication itself but, like authors, from OpenAlex's institution entities and then assigned to publications. All ROR IDs assigned by OpenAlex were verified using the ROR registry. They are correct and correspond to the institutions to which they are assigned.

In Scopus and WoS, the affiliation text (or 'address' in WoS) is complete and accurate in all cases. However, differences emerge in how each source processes and presents affiliation and institution information. OpenAlex and Scopus report affiliation strings exactly as they appear in publications, while WoS processes them to maintain a consistent structure. This structure starts with the main institution (e.g., university, hospital) and continues hierarchically with internal divisions of the institution (e.g., departments, centers). The names of institutions and divisions are also often abbreviated. The following example illustrates the differences between OpenAlex and Scopus on the one hand and WoS on the other hand:

- OpenAlex and Scopus: *Department of Mathematics, Faculty of Science and Technology, Women University of Azad Jammu and Kashmir, Bagh Pakistan*
- WoS: *Women Univ Azad Jammu & Kashmir, Fac Sci & Technol, Dept Math, Bagh, Pakistan*

In the institution field, unified institutions are also missing in WoS for nine institutions across six publications, similar to OpenAlex. In Scopus, all institutions are recorded, with only minimal errors (e.g., a city listed as part of the institution, or an institution split into two records). However, Scopus generally records more institutions per publication because it includes all hierarchical levels (department, faculty, university) as separate entries.

<u>Reference information</u>: As shown in Table 6, most publications (25 out of 30) in the sample are missing some references. One publication lacks a reference list entirely, while the others have incomplete ones. This is expected, as OpenAlex includes only linked references (i.e., those that refer to works also indexed in OpenAlex), rather than full bibliographies. Of the four publications without missing references, one is an editorial with no references. The remaining three have complete reference lists in OpenAlex.

On average, publications miss 30% of their references, considering those to both indexed and non-indexed works in OpenAlex. Most missing references are references to journal articles (45%), followed by references to books (20%), book chapters (9%), proceedings papers (7%), and reports (5.5%). Others include references to blogs, news articles, theses, datasets, and software. Following OpenAlex's approach to references, the works absent from reference lists should not be indexed in OpenAlex. However, manual checks indicate that 44% of the missing references are actually indexed in OpenAlex and are thus inaccurately absent in OpenAlex's reference lists. For references to journal articles, this percentage is even higher at 64%.

In Scopus and WoS, all publications have complete reference lists except for one missing reference to an R package in WoS. However, in some cases, reference information is incomplete (e.g., a missing title), but the cited work can in most cases be identified through other elements such as journal name, volume, issue, and page or article number.

Funding information: Nearly half of the publications (14 out of 30) contain funding acknowledgments. Of these, only seven list their complete list of funders in OpenAlex. Of the publications missing funders, four had no funders at all, and two had an incomplete list in which one of the funders was missing. Furthermore, all ROR IDs reported by OpenAlex were verified using the ROR registry and correspond correctly to the assigned institutions.

Scopus and WoS both record funding acknowledgment texts and identify funders mentioned in the texts. WoS provides two funder fields: funding agency and unified funding agency. The first provides the agency name as it appears in the acknowledgments text, while the second provides a standardized version. Out of 14 publications with funding acknowledgments,

Scopus is missing funding texts in five cases, and WoS in two. Regarding funders, Scopus is missing funders in six publications, and WoS in one case for the funding agency, and seven for the unified field.

Source information: Because journals were used to select the publications included in the analysis, the source field is not assessed for coverage, but only for accuracy. For all 30 publications, source information falls within the FI category. In some cases, discrepancies arise when comparing the information to that on the publication's landing page. However, these differences result from changes in the journals, such as title or ISSN changes. rather than errors in OpenAlex. All such cases are verified using the article PDFs and the ISSN Portal website.

Scopus and WoS also provide accurate source information for all publications, though Scopus omits four eISSNs and WoS omits one. All other information aligns with that available on the landing pages of the publications.

### *4.5 Metadata accuracy: publications only in OpenAlex*

As in the previous section, this section presents the results of a manual analysis of 30 randomly selected African publications indexed exclusively in OpenAlex (Subset 2), i.e., not indexed in Scopus or WoS. Table 7 summarizes the results of the metadata accuracy analysis, which are discussed in detail below.

**Table 7. Accuracy of metadata of African publications indexed only in OpenAlex based on a sample (n = 30) of publications**

| Metadata field | CMI | | PMI | | NMI | | FI | |
|---|---|---|---|---|---|---|---|---|
| | # | % | # | % | # | % | # | % |
| Publication information | | | | | | | | |
| DOI | 0 | 0 | - | - | 2 | 7 | 28 | 93 |
| Publication year | 0 | 0 | - | - | 4 | 12 | 26 | 88 |
| Volume | 7 | 23 | - | - | 1 | 3 | 22 | 73 |
| Issue | 6 | 20 | - | - | 2 | 7 | 22 | 73 |
| Article number | - | - | - | - | - | - | - | - |
| First page | 13 | 43 | - | - | 1 | 3 | 16 | 53 |
| Last page | 13 | 43 | - | - | 4 | 12 | 13 | 43 |
| OA status | 0 | 0 | - | - | 5 | 15 | 25 | 85 |
| Title | 0 | 0 | - | - | 0 | 0 | 30 | 100 |
| Abstract | 12 | 40 | - | - | 6 | 18 | 12 | 40 |
| Type | 0 | 0 | - | - | 1 | 3 | 29 | 97 |
| Language | 1 | 3 | - | - | 1 | 3 | 28 | 93 |
| Author information | | | | | | | | |
| Author names | 1 | 3 | 0 | 0 | 0 | 0 | 29 | 97 |
| Author order | 1 | 3 | 0 | 0 | 1 (2) | 3 | 29 | 97 |
| Corresponding author | 1 | 3 | 0 | 0 | 13 (13) | 43 | 16 | 53 |
| Affiliation and institution information | | | | | | | | |
| Raw affiliation strings | 17 | 57 | 0 | 0 | 6 (9) | 18 | 7 | 21 |
| Institutions | 23 | 80 | 1 (1) | 3 | 2 (2) | 7 | 4 | 12 |
| Reference information | | | | | | | | |
| References | 18 | 60 | 12 (249) | 40 | 3 (4) | 10 | 1 | 3 |
| Funding information | | | | | | | | |
| Funders | 3 | 10 | 0 | 0 | 0 | 0 | 27 | 90 |
| Source information | | | | | | | | |
| Journal title | 0 | 0 | 0 | 0 | 0 | 0 | 30 | 100 |
| e-ISSN | 0 | 0 | 0 | 0 | 0 | 0 | 30 | 100 |
| p-ISSN | 0 | 0 | 0 | 0 | 0 | 0 | 30 | 100 |

Publication information: Most fields related to publication information in OpenAlex are highly accurate in absolute terms. However, the DOI (93%) and publication year (88%) fields show lower FI values compared to Subset 1 (Table 6). These fields are particularly important, as they provide key information for bibliometric analyses and other uses.

Fields such as volume, issue, first page, and last page also exhibit discrepancies. In the case of the issue field, the FI percentage is higher than in Subset 1, largely because Subset 2 includes fewer journals that publish a single annual issue.

Accuracy for other fields is generally lower than in Subset 1. The abstract field, in particular, shows greater variability. 40% of publications are missing abstracts. Among the publications marked as NMI, three types of cases emerge. First, for most NMI cases, OpenAlex includes additional information in the abstract field, such as author names, journal titles, and keywords. Although the abstract text itself is otherwise complete and correct, this added content prompts classification as NMI because it may affect analyses based on abstracts. Second, in one case, the abstract is incomplete. Lastly, in one case, the abstract field contains a passage from the publication's full text instead of the abstract text itself.

Author information: OpenAlex accurately records the author names for 97% of the sample, a slight improvement over Subset 1. However, this result may reflect the characteristics of this specific sample and should not be generalized.

For ORCIDs, only one publication includes this identifier on its landing page or in the PDF, and OpenAlex correctly captures it. All other ORCIDs reported in OpenAlex were verified against the ORCID website and are accurate.

Author order is accurate in 97% of cases, with only a minor discrepancy noted. However, OpenAlex fails to identify the corresponding author for any multi-authored publications (43% of the sample), a notable difference compared to Subset 1.

Affiliation and institution information: Most publications in the sample lack complete affiliation and institution information, a striking difference compared to Subset 1. Specifically, 57% of publications are missing raw affiliation strings, and 80% are missing assigned institutions.

No cases of PMI are observed for raw affiliation strings, but six cases are classified as NMI. There are three reasons for NMI related to raw affiliation strings. First, in one case, OpenAlex reports an affiliation unrelated to the publication. Second, in another case, OpenAlex combines two affiliations into a single string, preserving the data but introducing a formatting issue that could affect downstream analyses. Lastly, in several cases, OpenAlex reports incomplete raw affiliation strings, omitting details such as the university, country, or department/unit. These cases are also categorized as NMI, as they could affect analyses that rely on this data.

Regarding institutions, one case of PMI occurs, where one institution is omitted. In the two NMI cases, OpenAlex incorrectly lists institutions unrelated to the publication. All ROR IDs assigned by OpenAlex were verified against the ROR registry and match the institutions to which they are assigned.

Reference information: The accuracy of reference information in Subset 2 is also notably lower compared to Subset 1. Due to overlapping issues of missing and erroneous information, some publications are classified into multiple categories. The majority of publications in the sample (60%) lack complete reference lists. Of the remaining publications, all but one (which has no references at all) include partial reference lists, missing between 6% and 95% of their references. The average share of missing references is 57%. Two NMI cases arise from references that are unrelated to the publications, and another involves a reporting error, where OpenAlex lists a book instead of a chapter from that book. Lastly, only one publication is classified as FI, and that publication is an index and introduction of a journal issue that includes no references.

An analysis of the missing references reveals that most refer to journal articles (34%), followed by books (30%), websites (11%, mostly from a single publication), and reports (5%). Other missing references include references to conference papers, chapters, datasets, software packages, theses, manuals, and surveys.

As noted in the previous section, following OpenAlex's approach to references, only references that refer to indexed works are expected to appear in the reference lists in OpenAlex. However, manual checks show that 39% of the missing references are actually indexed in OpenAlex and thus inaccurately absent.

Funding information: Most of the publications in the sample (90%) do not contain funding acknowledgements and therefore funding information is correctly absent in OpenAlex. However, for the three publications that do include funding acknowledgements, OpenAlex fails to report the mentioned funders. Since no funders are reported, ROR IDs for the funders are also missing.

Source information: OpenAlex provides complete and accurate source information for all 30 publications in the sample. In one instance, the journal title reported in OpenAlex differs from the version on the landing page of the publication due to a title change (similar to what was observed for some publications in Subset 1). All ISSNs, both print and electronic, match those on the landing pages and the ISSN Portal website.

## 5. Conclusions

This study shows that OpenAlex offers the most comprehensive coverage of the African publishing system among the four bibliographic data sources examined. It includes the majority of African research publications with a DOI indexed in Scopus, WoS, and AJOL, while also incorporating numerous publications that are not found in any of the other three data sources. In doing so, OpenAlex helps addressing the longstanding exclusion of African scholarship from mainstream bibliographic data sources, an issue widely recognized in the literature (Asubiaro, Onaolapo & Mills, 2024; Khana et al., 2022; Tijssen, 2007). Its broader coverage enhances the visibility of research outputs published outside dominant global publication outlets, thereby improving scientific diversity and reducing the disciplinary, linguistic, and regional gaps often found in traditional bibliometric sources (Alonso-Álvarez, 2024; Asubiaro & Onaolapo, 2023), ultimately offering a more complete perspective of the global scientific landscape.

When examining the metadata completeness of African research publications, OpenAlex demonstrates strong coverage in certain fields. It provides extensive coverage of publication and author information, especially for metadata fields such as issues, pages, generated keywords, and ORCIDs, where its coverage surpasses that of proprietary data sources. However, coverage of affiliations, references, and funder information is relatively lower. For publications also indexed in Scopus or WoS (Subset 1), metadata availability in OpenAlex is comparable to that of these proprietary data sources in most fields, including the majority of publication information, author, affiliation, reference, and funding fields. Conversely, for publications only indexed in OpenAlex (Subset 2), metadata coverage is generally more limited, particularly in critical fields such as affiliations, references, and funding.

In terms of metadata accuracy, the findings closely align with the patterns observed for completeness. Publications indexed in both OpenAlex and Scopus or WoS tend to exhibit higher accuracy across metadata fields. Publication information is generally accurate in both subsets, with pages and issues being the primary exceptions. The quality of OpenAlex metadata concerning abstracts, authors, institutions, references, and funding is lower than that of Scopus and WoS, particularly for publications exclusively indexed in OpenAlex. The results indicate that OpenAlex rarely records non-matching information, which is the most problematic type of error, as it leads to both omissions and false attributions. However, there are instances of partially missing information, and many publications in Subset 2 show completely missing metadata in several fields. Regarding references, it is important to note that while OpenAlex does not aim to provide complete reference lists, many references to indexed publications are also absent. Although prior studies have highlighted incomplete reference lists in OpenAlex (Alperin et al., 2024), the issue of missing citation links has received less attention. This limitation is critical, as it restricts the ability to trace citation patterns, particularly for publications not indexed in Scopus or WoS. Therefore, when aiming to conduct more inclusive citation analyses, it is important to acknowledge that results based on OpenAlex may still largely reflect citations from mainstream publication outlets, thereby potentially reproducing the biases inherent in traditional bibliographic data sources. Further research could investigate this issue more closely to better understand its causes and implications.

Several limitations should be considered when interpreting the results of this study. First, due to the process used to retrieve publications, which relies on a master list of sources, it was not possible to assess the coverage of metadata fields related to journal information. For instance, publications without ISSNs are excluded from the sample. Additionally, as OpenAlex is a resource that is rapidly evolving, the results of this study may not reflect its current state. This is due to both internal developments and improvements made by the OpenAlex team, as well as external decisions by other organizations. For example, in November 2024, Elsevier took steps to have abstracts of non-open access articles removed from OpenAlex, leading to the exclusion of approximately 11.5 million abstracts that were previously included in the database (Kramer, 2024).

When assessing the broader implications of the findings, several key conclusions emerge. First, OpenAlex shows significant potential as a replacement for traditional proprietary bibliographic data sources like Scopus and WoS, particularly for specific subsets of data. Notably, for publications also indexed in Scopus and WoS, the completeness and accuracy of metadata in OpenAlex are largely comparable, making it a viable option for many types of bibliometric studies. Moreover, its open nature, accessibility, and broad research coverage, including publications from underrepresented regions and disciplines, make it a valuable alternative, especially for researchers working in contexts where access to paid data sources is limited. However, while OpenAlex's efforts to enhance diversity and inclusion in scholarly coverage are commendable, the completeness and accuracy of its metadata still pose challenges in fields such as affiliations, references, and funding, which are essential for certain bibliometric studies. These limitations could impact analyses that rely on these fields, especially small-scale analyses or studies that depend on highly detailed information. This is particularly relevant for historically marginalized scientific communities, such as African journals and researchers, for whom the availability of robust, inclusive, and open bibliographic data sources is essential to improve both scholarly visibility and analytical representation.

Finally, the higher rates of missing data in some metadata fields compared to proprietary databases indicate that more work is needed. In this context, the scientific community must further engage with OpenAlex to better understand its limitations, identify areas for improvement, and expand its applicability for bibliometric and scientometric research. By systematically examining its current gaps, such as incomplete metadata fields or disparities in coverage across disciplines and regions, researchers can offer valuable feedback to enhance its coverage and metadata quality.

**Acknowledgements**

We thank the reviewers for their constructive feedback. Their suggestions and comments have significantly contributed to improving the quality of this manuscript.

**Open science practices**

The OpenAlex data (obtained from the OpenAlex snapshot) and AJOL data (retrieved from the AJOL website) used this paper is openly available. This is not the case for the data obtained from the Scopus and WoS databases. Due to license restrictions, we are not allowed to redistribute Scopus and WoS data, therefore the data used in this paper cannot be made available.

## Author contributions

Patricia Alonso-Álvarez: Conceptualization, Methodology, Formal analysis, Writing—original draft. Nees Jan van Eck: Conceptualization, Methodology, Writing— review & editing.

## Competing interests

The authors have no competing interests.

## Funding information

This study is funded under a PIPF contract of the Madrid Education, Science and Universities Office (reference: PIPF-2022/PH-HUM-25403). This work was also supported by the Spanish Ministry of Science under the project UnInCA (reference: PID2023-149340OB-I00).

## References

Alonso-Álvarez, P. (2024). Exploring research quality and journal representation: A comparative study of African Journals Online, Scopus, and Web of Science. *Research Evaluation*, *33*, rvae057. https://doi.org/10.1093/reseval/rvae057

Amboka, P., Sindi, J. K., Wamukoya, M., Orobaton, N., Neba, A., Vicente-Crespo, M., & Gitau, E. (2024). Discoverability of African journals by Google scholar and inclusion in Scopus. *VeriXiv*, *1*(17), 17. https://verixiv.org/articles/1-17/v1?src=rss

Alperin, J. P., Portenoy, J., Demes, K., Larivière, V., & Haustein, S. (2024). *An analysis of the suitability of OpenAlex for bibliometric analyses* (arXiv:2404.17663). arXiv. https://doi.org/10.48550/arXiv.2404.17663

Archambault, É., Vignola-Gagné, É., Côté, G., Larivi?re, V., & Gingrasb, Y. (2006). Benchmarking scientific output in the social sciences and humanities: The limits of existing databases. *Scientometrics*, *68*(3), 329-342. https://doi.org/10.1007/s11192-006-0115-z

Arroyo-Machado, W., & Costas, R. (2023). Do popular research topics attract the most social attention? A first proposal based on OpenAlex and Wikipedia. *Proceeding of the 27th International Conference on Science, Technology and Innovation Indicators (STI 2023)*. https://doi.org/10.55835/6442bb04903ef57acd6dab9e

Asubiaro, T., Onaolapo, S., & Mills, D. (2024). Regional disparities in Web of Science and Scopus journal coverage. *Scientometrics*, *129*(3), 1469-1491. https://doi.org/10.1007/s11192-024-04948-x

Asubiaro, T. V., & Onaolapo, S. (2023). A comparative study of the coverage of African journals in Web of Science, Scopus, and CrossRef. *Journal of the Association for Information Science and Technology*, *74*(7), 745-758. https://doi.org/10.1002/asi.24758

Barcelona Declaration on Open Research Information (2024). Barcelona Declaration on Open Research Information. *Zenodo*. https://doi.org/10.5281/zenodo.10958522

Becerra, G. (2024). *ojsr: Crawler and Data Scraper for Open Journal System ('OJS')* (Versión 0.1.5) [Software]. https://cran.r-project.org/web/packages/ojsr/index.html

Boshoff, N., Ngwenya, S., Uisso, A. J., Koch, S., Costas, R., & Dudek, J. (2024, November 15). *Different representations of forest science in bibliographic databases and the (in-)visibility of Tanzanian research: Applying an epistemic (in-)justice lens*. Zenodo. https://doi.org/10.5281/zenodo.14171522

Bratt, S., Langalia, M., & Nanoti, A. (2023). North-south scientific collaborations on research datasets: A longitudinal analysis of the division of labor on genomic datasets (1992–2021). *Frontiers in Big Data*, *6*. https://doi.org/10.3389/fdata.2023.1054655

Castro Torres, A. F. (2024). Revisiting North-South Disparities in Naming Practices: An Extension and Update. *Social Currents*, 23294965241297932. https://doi.org/10.1177/23294965241297932

Céspedes, L., Kozlowski, D., Pradier, C., Sainte-Marie, M. H., Shokida, N. S., Benz, P., ... & Larivière, V. (2025). Evaluating the linguistic coverage of OpenAlex: An assessment of metadata accuracy and completeness. *Journal of the Association for Information Science and Technology*. https://doi.org/10.1002/asi.24979

Chavarro, D., & Alperín, J. P. (2024). Equity in Scholarly Visibility: Bridging the Gap for Journals using Open Journal Systems in OpenAlex. *Proceeding of the 28th International Conference on Science, Technology and Innovation Indicators (STI 2024)* https://doi.org/10.5281/zenodo.13951504

Chavarro, D., Ràfols, I., & Tang, P. (2018). To what extent is inclusion in the Web of Science an indicator of journal 'quality'? *Research Evaluation*, *27*(2), 106-118. https://doi.org/10.1093/reseval/rvy001

Ciarli, T., & Ràfols, I. (2019). The relation between research priorities and societal demands: The case of rice. *Research Policy*, *48*(4), 949-967. https://doi.org/10.1016/j.respol.2018.10.027

Culbert, J. H., Hobert, A., Jahn, N., Haupka, N., Schmidt, M., Donner, P., & Mayr, P. (2025). Reference coverage analysis of OpenAlex compared to Web of Science and Scopus. *Scientometrics, 130*(4), 2475-2492. https://doi.org/10.1007/s11192-025-05293-3

Curry, M. J., & Lillis, T. M. (2010). *Academic Writing in a Global Context: The Politics and Practices of Publishing in English*. Routledge.

Delgado-Quirós, L., & Ortega, J. L. (2024). Completeness degree of publication metadata in eight free-access scholarly databases. *Quantitative Science Studies*, *5*(1), 31-49. https://doi.org/10.1162/qss_a_00286

Haunschild, R., & Bornmann, L. (2024). The use of OpenAlex to produce meaningful bibliometric global overlay maps of science on the individual, institutional, and national levels. *PloS one*, 19(12), e0308041. https://doi.org/10.1371/journal.pone.0308041

Haupka, N., Culbert, J. H., Schniedermann, A., Jahn, N., & Mayr, P. (2024). *Analysis of the Publication and Document Types in OpenAlex, Web of Science, Scopus, Pubmed and Semantic Scholar* (arXiv:2406.15154). arXiv. https://doi.org/10.48550/arXiv.2406.15154

Hountondji, P. (1990). Scientific Dependence in Africa Today. *Research in African Literatures*, *21*(3), 5–15. https://www.jstor.org/stable/3819631

Jahn, N., Haupka, N., & Hobert, A. (2023). *Scholarly Communication Analytics: Analysing and reclassifying open access information in OpenAlex*. https://subugoe.github.io/scholcomm_analytics/posts/oalex_oa_status/

Khanna, S., Ball, J., Alperin, J. P., & Willinsky, J. (2022). Recalibrating the scope of scholarly publishing: A modest step in a vast decolonization process. *Quantitative Science Studies*, *3*(4), 912-930. https://doi.org/10.1162/qss_a_00228

Klebel, T., & Ross-Hellauer, T. (2023). The APC-barrier and its effect on stratification in open access publishing. *Quantitative Science Studies*, *4*(1), 22-43. https://doi.org/10.1162/qss_a_00245

Kramer, B. (2024). More open abstracts? Comparing abstract coverage in Crossref and OpenAlex. *Sesame Open Science*. https://bmkramer.github.io/SesameOpenScience_site/thought/202411_open_abstracts/

Kumar, A., Koley, M., Yegros, A., & Rafols, I. (2024). Priorities of health research in India: Evidence of misalignment between research outputs and disease burden. *Scientometrics*, *129*(4), 2433-2450. https://doi.org/10.1007/s11192-024-04980-x

Larivière, V., Haustein, S., & Mongeon, P. (2015). The Oligopoly of Academic Publishers in the Digital Era. *PLOS ONE*, *10*(6), e0127502. https://doi.org/10.1371/journal.pone.0127502

Mazzoni, A., & Costas, R. (2024, junio 6). Towards the democratisation of open research information for scientometrics and science policy: The Campinas experience. *Leiden Madtrics*. https://www.leidenmadtrics.nl/articles/towards-the-democratisation-of-open-research-information-for-scientometrics-and-science-policy-the-campinas-experience

Mills, D., Kingori, P., Branford, A., Tamti Chatio, S., Robinson, N., & Tindana, P. (2023). *Who Counts? Ghanaian Academic Publishing and Global Science*. African Minds. https://www.africanminds.co.za/who-counts/

Mongeon, P., Bowman, T. D., & Costas, R. (2023). An open data set of scholars on Twitter. *Quantitative Science Studies*, *4*(2), 314-324. https://doi.org/10.1162/qss_a_00250

Mongeon, P., & Paul-Hus, A. (2016). The journal coverage of Web of Science and Scopus: A comparative analysis. *Scientometrics*, *106*(1), 213-228. https://doi.org/10.1007/s11192-015-1765-5

Moscona, J., & Sastry, K. (2022). *Inappropriate Technology: Evidence from Global Agriculture* (SSRN Scholarly Paper 3886019). Social Science Research Network. https://doi.org/10.2139/ssrn.3886019

Okamura, K. (2024). *Evolving interdisciplinary contributions to global societal challenges: A 50-year overview* (arXiv:2410.20619). arXiv. https://doi.org/10.48550/arXiv.2410.20619

Ogunfolaji, O., Tangmi, A., Dada, O. E., Sebopelo, L. A., Sichimba, D., Djoutsop, O. M., ... & Esene, I. (2022). Profiling African Health Journals: A Bibliometric Study. *International Journal of Public Health*, *67*, 1604932. https://doi.org/10.3389/ijph.2022.1604932

Perianes-Rodriguez, A., Gomez-Nuñez, A. J., & Olmeda-Gomez, C. (2024). Anatomy of the top 1% most highly cited publications: An empirical comparison of two approaches. *Quantitative Science Studies*, *5*(2), 447-463. https://doi.org/10.1162/qss_a_00290

Priem, J., Piwowar, H., & Orr, R. (2022). *OpenAlex: A fully-open index of scholarly works, authors, venues, institutions, and concepts* (arXiv:2205.01833). arXiv. https://doi.org/10.48550/arXiv.2205.01833

Schares, E. (2024). Comparing Funder Metadata in OpenAlex and Dimensions. *OpenISU*. https://doi.org/10.31274/b8136f97.ccc3dae4

Simard, M.-A., Basson, I., Hare, M., Lariviere, V., & Mongeon, P. (2024). *The open access coverage of OpenAlex, Scopus and Web of Science* (arXiv:2404.01985). arXiv. https://doi.org/10.48550/arXiv.2404.01985

Singh, P., & Singh, V. K. (2023). Exploring the publication metadata fields in Web of Science, Scopus and Dimensions: Possibilities and ease of doing scientometric analysis. *Proceedings of ISSI 2023— the 19th International Conference of the International*

*Society for Scientometrics and Informetrics*, *1*, 579-601. https://doi.org/10.5281/zenodo.8306017

Sorbonne University. (2023). Sorbonne University unsubscribes from the Web of Science. https://www.sorbonne-universite.fr/en/news/sorbonne-university-unsubscribes-web-science

Tijssen, R. J. W. (2007). Africa's contribution to the worldwide research literature: New analytical perspectives, trends, and performance indicators. *Scientometrics*, *71*(2), 303-327. https://doi.org/10.1007/s11192-007-1658-3

Van Eck, N.J., Waltman, L., & Neijssel, M. (2024). Launch of the CWTS Leiden Ranking Open Edition 2024. *Leiden Madtrics*. https://doi.org/10.59350/r512t-r8h93

Van Leeuwen, T. N., Moed, H. F., Tijssen, R. J. W., Visser, M. S., & Van Raan, A. F. J. (2001). Language biases in the coverage of the Science Citation Index and its consequencesfor international comparisons of national research performance. *Scientometrics*, *51*(1), 335-346. https://doi.org/10.1023/A:1010549719484

Velez-Estevez, A., Perez, I. J., García-Sánchez, P., Moral-Munoz, J. A., & Cobo, M. J. (2023). New trends in bibliometric APIs: A comparative analysis. *Information Processing & Management*, *60*(4), 103385. https://doi.org/10.1016/j.ipm.2023.103385

Vera-Baceta, M.-A., Thelwall, M., & Kousha, K. (2019). Web of Science and Scopus language coverage. *Scientometrics*, *121*(3), 1803-1813. https://doi.org/10.1007/s11192-019-03264-z

Waltman, L., Van Eck, N.J., Visser, M., Neijssel, M., Montgomery, L., Neylon, C., ... & Priem, J. (2024). Introducing the Leiden ranking open edition. Leiden Metrics. *Leiden Madtrics*. https://doi.org/10.59350/89wpz-hpz32

Wickham, H., Software, P., & PBC. (2024). *rvest: Easily Harvest (Scrape) Web Pages* (Versión 1.0.4) [Software]. https://cran.r-project.org/web/packages/rvest/index.html

Zhang, L., Cao, Z., Shang, Y., Sivertsen, G., & Huang, Y. (2024). Missing institutions in OpenAlex: Possible reasons, implications, and solutions. *Scientometrics*. https://doi.org/10.1007/s11192-023-04923-y